\documentclass[a4paper,12pt,epsfig]{article}
\usepackage{epsfig}
\usepackage{epsfig}
\usepackage{amssymb}
\usepackage{amsfonts}

\usepackage{amsmath}
\newskip\humongous \humongous=0pt plus 1000pt minus 1000pt

\newif\ifdtup

\catcode`\@=11

\@addtoreset{equation}{section}
\def\theequation{\thesection\arabic{equation}}

\def\@normalsize{\@setsize\normalsize{15pt}\xiipt\@xiipt
\abovedisplayskip 14pt plus3pt minus3pt%
\belowdisplayskip \abovedisplayskip
\abovedisplayshortskip \z@ plus3pt%
\belowdisplayshortskip 7pt plus3.5pt minus0pt}

\def\small{\@setsize\small{13.6pt}\xipt\@xipt
\abovedisplayskip 13pt plus3pt minus3pt%
\belowdisplayskip \abovedisplayskip
\abovedisplayshortskip \z@ plus3pt%
\belowdisplayshortskip 7pt plus3.5pt minus0pt
\def\@listi{\parsep 4.5pt plus 2pt minus 1pt
\itemsep \parsep
\topsep 9pt plus 3pt minus 3pt}}

\relax

\catcode`@=12

\catcode`\@=11

\def\section{\@startsection{section}{1}{\z@}{3.5ex plus 1ex minus
.2ex}{2.3ex plus .2ex}{\large\bf}}

\def\thesection{\arabic{section}.}

\def\appendix{\setcounter{section}{0}
\def\thesection{Appendix \Alph{section}:}
\def\theequation{\Alph{section}.\arabic{equation}}}

\newcommand{\be}{\begin{equation}}
\newcommand{\ee}{\end{equation}}
\newcommand{\bqa}{\begin{eqnarray}}
\newcommand{\eea}{\end{eqnarray}}
\newcommand{\beas}{\begin{eqnarray*}}
\newcommand{\eeas}{\end{eqnarray*}}

\newcommand{\non}{\nonumber}
\newcommand{\bquo}{\begin{quote}}
\newcommand{\enqu}{\end{quote}}

\def\Tr{ \hbox{\rm Tr}}

\def\brc{\langle}
\def\ckt{\rangle}

\def\diag{\hbox{\rm diag}}

\begin{document}

\begin{titlepage}
\begin{flushright}
IFUP-TH/2005-05\\
hep-th/0502004\\
\end{flushright}
\bigskip
\bigskip
\def\thefootnote{\fnsymbol{footnote}}

\begin{center}
{\large  {\bf
Light Nonabelian Monopoles and Generalized $r$-Vacua in Supersymmetric Gauge Theories
 } }
\end{center}

\renewcommand{\thefootnote}{\fnsymbol{footnote}}
\bigskip
\begin{center}
{\large  Stefano BOLOGNESI\footnote{s.bolognesi@sns.it} $^{(1,3)}$,
 Kenichi KONISHI\footnote{konishi@mail.df.unipi.it} $^{(2,3)}$,  \\  Giacomo
 MARMORINI\footnote{g.marmorini@sns.it} $^{(1,3)}$}
 \vskip 0.10cm

\end{center}

\renewcommand{\thefootnote}{\arabic{footnote}}
\setcounter{footnote}{0}

\begin{center}
{\it      \footnotesize
Scuola Normale Superiore - Pisa $^{(1)}$,
 Piazza dei Cavalieri 7, Pisa, Italy \\
Dipartimento di Fisica ``E. Fermi" -- Universit\`a di Pisa $^{(2)}$, \\
Istituto Nazionale di Fisica Nucleare -- Sezione di Pisa $^{(3)}$, \\
Largo Pontecorvo, 3, Ed. C, 56127 Pisa,  Italy,
}

\end {center}

\bigskip
\bigskip

\noindent
\begin{center} {\bf Abstract} \end{center}
We study a class of ${\cal N}=1 $ supersymmetric $U(N)$ gauge
theories and find that there exist vacua in which the low-energy
magnetic effective gauge group contains multiple nonabelian
factors, $\prod_{i} SU(r_{i})$,  supported by light monopoles
carrying the associated  nonabelian charges. These nontrivially
generalize the physics of the so-called $r$-vacua found in softly broken ${\cal
N}=2$  supersymmetric $SU(N)$ QCD, with  an  effective
low-energy gauge group $SU(r) \times U(1)^{N-r}$.
  The  matching between classical and quantum $(r_{1}, r_{2},\ldots)$
vacua  gives an  interesting hint about the nonabelian duality.

\vfill

\begin{flushright}
\today
\end{flushright}
\end{titlepage}

\section{Introduction}

The quantum  behavior of  nonabelian monopoles  in  spontaneously
broken nonabelian gauge systems  is of considerable  interest. It
could for instance be a key for understanding the confinement in
QCD. In general, semiclassical ``nonabelian monopoles''   can
either disappear,  leaving only abelian monopoles detectable in
low-energy theory, or   survive as weakly coupled low-energy
degrees of freedom, with genuine magnetic gauge interactions.
Still another possibility is that the theory flows into a
nontrivial conformal theory  in which abelian or nonabelian
monopole fields appear together with relatively nonlocal dyons and
quarks. All of these  possibilities are realized in various vacua
of  softly broken ${\cal N}=2$ supersymmetric theories or in a
large class of ${\cal N}=1$ gauge theories. A point of fundamental
importance \cite{ABEKM} is that nonabelian monopoles \footnote{For
classical treatments and more recent developments on these, see
\cite{NAmonop}}, in contrast to abelian ones, are essentially
quantum mechanical.  Often,  a reasoning relying  on the
semiclassical  approximation only gives an incorrect picture of
the full quantum behavior.   With the help of certain exact
knowledge about the dynamical properties of supersymmetric gauge
theories, we have now acquired considerable control on the quantum
behavior of light nonabelian monopoles. Let us summarize the
situation:
\begin{description}
  \item  [(i)]  In pure ${\cal N}=2$  gauge theories, softly broken   to ${\cal N }=1$ by the  adjoint scalar mass only,
  the low-energy  theory is an effective magnetic abelian $U(1)^{R}$ gauge theory, where $R$ is the rank of the group;
  all monopoles are abelian  and the theory effectively  abelianizes \cite{SW1,curves,DS};
  \item[(ii)]  In ${\cal N}=2$  gauge theories  with $N_{f}$  flavors (fields in the fundamental representation of the gauge group),
  many vacua  exist where the low-energy effective magnetic gauge symmetry is nonabelian, $SU(r) \times U(1)^{R-r}$, $R$ being the rank of the group, $r= 1,2, \ldots,
\frac{N_{f}}{ 2}$.
 The light degrees of freedom in these vacua are  nonabelian monopoles  transforming as ${\underline r}$  of $SU(r)$ and carrying one of the $U(1)$ charges;   abelian
 monopoles having charge in different $U(1)^{R-r-1}$ groups appear as well.
 Upon   ${\cal N }=1$ perturbation, these monopoles condense and give rise to confinement (nonabelian dual superconductor) \cite{SW2,APS,CKM,BK};
  \item[(iii)]  Abelian and nonabelian Argyres-Douglas  vacua in   ${\cal N}=2$ theories \cite{SCFT}:  abelian or nonabelian monopoles and dyons together appear  at a nontrivial infrared  fixed point theory.
With a soft ${\cal N}=1$ perturbation, such vacua confine;

\item[(iv)] Many ${\cal N }=1$ conformal theories are known, having  dual descriptions \`a la Seiberg,  Kutasov, Kutasov-Schwimmer, etc. \cite{Sei,KS,KSS}

\end{description}

In this paper we continue our investigation and in particular,
study  systems in which at high energies the theory is a $U(N)$
gauge theory with an adjoint chiral multiplet $\Phi$ and  a set of
quark multiplets $Q$, ${\widetilde Q}$  with some superpotential,
while the low-energy magnetic gauge group  contains more than one
nonabelian factor, {\it e.g.},    $SU(r_{1}) \times SU(r_{2})
\times  SU(r_{2})  \ldots$, supported by light  monopoles, so that
they do not abelianize dynamically. The key ingredient  turns out
to be the superpotential,
\be   {\cal W}(\Phi, Q, {\widetilde Q})    =    W (\Phi)  +  {\widetilde Q} \,  m (\Phi) \, Q,
\label{superp}\ee
with a nontrivial structure in the function  $m(\Phi)$.

\section {Bosonic $SU(N)$ theory with an adjoint scalar \label{sec:ordinary}}

As a way of illustration  let us consider first  a bosonic  $SU(N
+ 1 )$ model \be {\cal L}= \frac{ 1}{ 4 g^2} (F_{\mu \nu}^A)^2 + \frac{ 1}{
g^2} |({\cal D}_{\mu} \phi)^A|^2 - V(\phi), \ee where $\phi$ is  a
complex scalar field in the adjoint representation of $SU(N+1)$.
Suppose  that the potential is such that at a minimum an adjoint
scalar VEV takes the form \be \brc \phi \ckt =
\begin{pmatrix}
    v_1 \cdot {\bf 1}_{r_1 \times r_1}    &   &  \\
     & v_2 \cdot {\bf 1}_{r_2 \times r_2}  &  \\
     &  & \ddots
\end{pmatrix}
\label{phivev} \ee
where  other diagonal elements, which we take
all different, are represented by dots. Such a VEV breaks the
gauge symmetry as \be SU(N) \to \frac{SU(r_1 ) \times SU(r_2 ) \times
U(1)^{N-r_1 -r_2 +1  }   }{ {\mathbb Z}_{r_1} \times {\mathbb
Z}_{r_2} }. \label{Symbr}\ee This system possesses several
semiclassical  monopoles. As is well known \cite {NAmonop,ABEKM}   they lie
in various broken $SU(2)$ subgroups,
\begin{description}
\item{i)} in $(i, N+1)$ subspaces, $i=1,2,\ldots, r_1$, giving rise to independent set of $r_1$ degenerate monopoles. These are
the semiclassical candidates for the nonabelian monopoles in ${\underline r}_{1}$ of the  dual $SU(r_{1})$ group;
\item{ii)} in $(j, N+1)$ subspaces, $j=r_1+1, r_1+2,\ldots, r_1+ r_2$, giving rise to $r_2$ degenerate monopoles, possibly in
${\underline r}_{2}$ of the  dual $SU(r_{2})$ group;
\item{iii)} in $(i,j)$ subspaces $i=1,2,\ldots, r_1$, $j=r_1+1, r_1+2,\ldots, r_1+ r_2$, giving rise to $r_1 \, r_2$
degenerate monopoles. These could be components of the nonabelian monopoles in $({\underline r}_{1}, {\underline r}_{2})$
representation of the dual  $SU(r_{1})\times SU(r_{2})$ group.
\end{description}
With appropriately chosen $v_1$, $v_2$, it is easy to arrange
things so that monopoles of the third type  are the lightest of
all. We shall find that such semiclassical reasoning however does
not give a correct picture of the quantum theory, as $v_1$ and
$v_2$ are taken small. Also, we would like to know whether there
are vacua in which  these monopoles appear as IR degrees of
freedom and play the role of  order parameters of confinement, and
find out in which type of theories this occurs.

The question is highly nontrivial: for instance,   in  softly
broken  ${\cal N}=2$ SQCD with superpotential
\be   {\cal   W} =  \sqrt{2} \, {\widetilde Q}_i  \, \Phi \, Q_i + \mu \, \Tr \,\Phi^2 +  m_i \, {\widetilde Q}_i  \, Q_i, \qquad  m_i \to 0,
\,\,   \mu \ll \Lambda,   \ee
 we know that   possible quantum vacua in the limit $m_{i} \to 0$ can be completely classified \cite{CKM} by an integer $r$.  The low-energy effective
magnetic gauge symmetry is in general nonabelian and of type
 \be  SU(r) \times U(1)^{N_{c}-r +1},  \qquad r=0,1,2,\ldots, \frac {N_f}{2}.
 \ee
In other words,  vacua with low-energy gauge symmetry of the type
Eq.(\ref{Symbr}) do not occur,  even though this fact is not
obvious from semiclassical reasoning.
 We are interested in knowing whether there exist  systems in which an effective gauge symmetry with multiple nonabelian factors
 occur at low energies.

\section{Generalized $r$-Vacua  in  ${\cal N}=1$    $U(N)$  Theories: Semi-Classical Approximation
\label{sec:semic}}

With the purpose of finding these new types of vacua, we enlarge
the class of theories, and consider ${\cal N}=1$  supersymmetric
$U(N)$ gauge theories with a matter field  $\Phi$ in the adjoint
representation and a set of quark superfields $Q_{i}$ and
${\widetilde Q}^{i}$ in the fundamental (antifundamental)
representation of $U(N)$, with a more general class of
superpotentials:
  \be  {\cal W} = W( \Phi )  +  \,{\widetilde Q}_i^a \,  m_{i}(\Phi)_a^b \,  Q_b^i:
\label{theory}    \ee $i=1,2,\ldots N_f $ is the flavor index;
$a,b=1,2,\ldots N$ are the color indices. In the flavor-symmetric
limit   $ m_{i}(\Phi)  \to    m(\Phi) $ (independent of  $i$) the
theory is invariant under  a global $U(N_f)$ symmetry.

Apart from the simplest cases (up to cubic functions  $ W( \Phi )
$ and  up to linear function  $ m(\Phi) $) the models considered
are not renormalizable.  As explained  in Ref.\cite{KSS}, however,
these potentials represent ``dangerously irrelevant''
perturbations,  and cannot be neglected in understanding the
dynamical behavior of the theory in the infrared.  We consider
Eq.(\ref{theory}) as an effective Lagrangian at a given scale, and
then explore the properties  of the theory as the  mass scale is
reduced towards zero.

The semiclassical vacuum equations are
\bqa &&    [ \Phi, \Phi^{\dagger}] = 0 \, ;
\label{D1}
\\   &&   0  = Q_a^i (Q^{\dagger})_i^b - ({\widetilde Q}^{\dagger})_a^i
{\widetilde Q}_i^b \, ; \label{D2}
\\
&&   Q_a^i  \, \frac{\delta m_i(\Phi)_{ a}^{b}  }{ \delta  \Phi_{c}^{d}}  \,   {\widetilde Q}_i^b     + \frac {\delta W(\Phi) }{  \delta \Phi_{c}^{d}}   = 0 \, ;
\label{F1}
\\
&&   m_i(\Phi)_{a}^b Q_b^i = 0 \qquad ( { \hbox {\rm no sum
over}} \, \,i) \, ;
\label{F2}
\\
&&   {\widetilde Q}_i^b \,m_i(\Phi)_{b}^a= 0 \qquad ({\hbox{\rm no sum over}} \,\,i).
\label{F3} \eea
 As explained  in \cite{CKM}, it is convenient first
to consider flavor nonsymmetric cases, {\it i.e.}, generic
$m_{i}(\Phi)$ and nonvanishing $ {\cal W}$ , so that the only
vacuum degeneracy  left is a discrete one, and to take the
$U(N_{f})$ limit  $m_{i}(\Phi) \to m(\Phi)$  only after
identifying each vacuum  and computing the condensates in it. $
m_{i}(\Phi) $ and $W(\Phi)$  are taken in  the general form
\be   W(\Phi) =  \sum_{k} \,a_{k}\, \Tr \, (\Phi^{k}), \qquad [ m_{i}(\Phi) ]_{ab}     = \sum_k    m_{i,k}  \Phi^{k-1}_{ab},
 \ee
furthermore they are assumed to satisfy no special relations.   We
shall choose  $m(\Phi)$  to be a polynomial of order quadratic or
higher, and assume that the equation in the flavor symmetric limit
\be  m(z)=0
\ee
has  distinct  roots
\be    z=v^{(1)}, v^{(2)}, v^{(3)}, \ldots.
\ee
Just as a reference,  in the ${\cal N}=2 $ theory broken to ${\cal
N}=1$  only by the adjoint mass term, $W(\Phi) = \mu \, \Tr \,
\Phi^{2},$    $m_{i}(\Phi) = \sqrt{2} \Phi  + m_{i}$, so that in
that case  the  flavor symmetric equation $m(z)=0$  would have a
unique root, $-  m/\sqrt{2}.$

We  first use a gauge  $U(N)$ rotation to bring the $\Phi$ VEV
into a diagonal form,
\be
\Phi =\delta_{ab} \,  \phi_{a} =  \diag \, (\, \phi_1, \phi_2, \ldots \phi_{N}\, ),
\label{phivevv}
\ee
which solves Eq.(\ref{D1}).
$m(\Phi)$  is then also diagonal in color,
\be [ m_{i}(\Phi) ]_{ab}    =  \sum_k    m_{i,k}  \delta_{ab} \phi_a^{k-1}   =  \delta_{ab} \, m_{i}(\phi_a), \quad i=1,2,\ldots, N_{f}.
\ee
 Eqs.(\ref{F1}),(\ref{F2}),(\ref{F3})   become
\bqa &&  \sum_{i }\, m_{i}^{\prime}(\phi_c)  \, Q_c^{i} \,{\widetilde Q}_i^{c} +  W^{\prime}(\phi_c) =0;  \label{quarkc}
\\
 &&  m_{i}(\phi_c)   \, Q_c^{i} =0; \qquad    m_{i}(\phi_c)  \, {\widetilde Q}_i^{c} =0;   \label{phic}
\eea
(no sum over $c$), and
\be     Q_c^{i} \,[ \,  \sum_{k}  m_{i,k} \, \sum_{\ell}  \,\phi_{c}^{\ell}  \, \phi_{d}^{k-\ell-1}\,] \,  {\widetilde Q}_i^{d}=0, \quad (c\ne d).
\label{nondVEVs} \ee

The diagonal elements   $\phi_c$  (and the squark condensate) are
of two different types.  The first corresponds to  one of the
roots of
\be  m_{i}(\phi_c^{*}) =0.   \label{sold}
\ee
For each $c$  this equation can be satisfied at most for one
flavor, as we consider the generic, unequal functions
$m_{i}(\Phi)$ first. Then there is one squark pair $ Q_c^{i},\,
{\widetilde Q}_i^{c} $ with nonvanishing VEVs and
Eq.(\ref{quarkc})  yields their VEVs
\be
Q_c^{i} = {\widetilde Q}_i^{c} =  \sqrt{-  \frac{ W^{\prime}(\phi_c^*)
}{ m_{i}^{\prime}(\phi_c^*)  } } \ne 0.
\ee
Note that according to our assumption  $m$ and $W$ satisfy  no
special relations  so that $W^{\prime}(\phi_c^*) \ne 0$.

The second group  of $\brc \phi_{c} \ckt $   corresponds to:
\be       W^{\prime}(a_{j}) =0, \qquad j=1,2,\ldots.
\ee
As  in general $m_{i}(a_{j})\ne 0$, we have
\be   Q_c^{i} = {\widetilde Q}_i^{c} =  0,
\ee
for the corresponding color components of the squarks.

The classification of the vacua  is somewhat subtle.
The adjoint scalar VEV  has  the form
 \be
\Phi = \diag \, (\underbrace { {v_{1}^{(1)}, v_{2}^{(1)}, \ldots, v_{r_{1}}^{(1)} }}_{r_{1}},
\ldots,
   {\underbrace { { v_{1}^{(p)}, \ldots , v_{r_{1}+\ldots  +r_{p}}^{(p)} }}_{r_{p}}},
    \ldots,  a_{1},   \ldots,  a_{n} ),
\label{phiVEVs}
\ee
namely, there appear $r_{1}$ roots   near $v^{(1)},$\,  $r_{2}$ roots   near $v^{(2)},$  and so on.

The diagonal elements ($ a_{1},  \ldots,  a_{n} $)  in
Eq.(\ref{phiVEVs})  correspond to the roots of
$W^{\prime}(z)=0$:   these color components  give rise to pure
$U(N_{1}) \times U(N_{2}) \times \ldots U(N_{n})$ theory,  $\sum
N_{i} =  N -  \sum r_{j},$  where   $N_{j}$ corresponds to the
number of times $a_{j}$ appears in Eq.(\ref{phiVEVs}).    No
massless matter charged with respect to $SU(N_{j})$ group are
there, $\prod SU(N_{j})$ interactions become strong in the
infrared, and yields an  abelian  $U(1)^{n}$ theory. The physics
related to this part of the system has been recently discussed
extensively, by use of a Matrix model conjecture proposed by
Dijkgraaf-Vafa \cite{DV} as well as by  a  field theory approach
initiated by Cachazo-Douglas-Seiberg-Witten \cite{CDSW}.  We have
nothing to add to them here.

Our focus of attention  here is complementary:   it  concerns the
first $r_{1}+ r_{2}+\ldots + r_{p}$ color components of $\Phi$ in
Eq.(\ref{phiVEVs}).  This is interesting because in the flavor
symmetric and     $W(\Phi) \to 0$  limit,    this sector describes
a  local
\be  U(r_{1})\times  U(r_{2})\times \ldots U(r_{p}) \label{magnetic}
\ee
gauge theory, supported by $N_{f}$ massless quarks in the
fundamental representation of $SU(r_{1})$, $N_{f}$ massless quarks
in the fundamental representation of $SU(r_{2})$, and so on.  If
$r_{i} \le \frac{N_{f}}{2}$ these interactions are non
asymptotically free\footnote{Note that   the  mass terms come from
$\theta \theta$ component  of
\be      \,{\widetilde
Q}_i^a  m_{i}(\Phi)_a^b Q_b^i   \sim     ({\widetilde Q} +  \theta
\, {\widetilde \psi_Q}  ) \, ( 0 +\theta \, m^{\prime}(v_1) \,
\psi ) \,   ({ Q} +  \theta \, {\psi}_Q  )  + \ldots
\ee
so in the limit $Q_i \to 0$ no mass terms arise  and the beta
function of   ${\cal N}=2$  theory can be used even if interaction
terms conserve the  ${\cal N}=1 $ supersymmetry only. } and   they
remain weakly coupled (or at most evolve to a superconformal
theory) in the low energies. Furthermore, if all condensates are
small or of order of $\Lambda$,  Eq.(\ref{magnetic}) describes a
magnetic theory, the breaking of $U(r_{i})$  by nonzero $W(\Phi)$
is a nonabelian Meissner effect (nonabelian confinement).

As for the structure of squark condensates,
Eq.(\ref{quarkc})-Eq.(\ref{nondVEVs})  appear to imply  a
color-flavor locked form of the squark VEVs  of the form
\be     Q_c^{i}\propto  \delta_{c}^{i}, \quad  {\widetilde Q}_i^{c}  \propto  \delta_{i}^{c}.
\ee
This  however turns out to be  true only if the groups of $r_{1}$
elements $v_{i}^{(1)}$,  $r_{2}$  elements $v_{i}^{(2)}$, etc,
correspond to zeros of {\it mutually exclusive sets }  of flavor
functions $m_{i}(z)$,  which is not necessarily the case.

Let us explain this point  better.  Suppose that the two of the
diagonal $\phi$  VEVs in Eq.(\ref{phiVEVs}), $x_{1} \equiv
v_{1}^{(1)} $ and  $x_{2} \equiv v_{r_{1}+1}^{(2)}$, correspond
both to the first flavor.  The VEVs of the first quark has  the
form {\small
 \be  Q^{1}= \left(\begin{array}{c}d_1    \\ 0     \\ \vdots    \\ 0 \\
  d_{1}^{\prime} \\ 0 \\ \vdots \\ 0
    \end{array}\right), \quad   {\widetilde Q}_{1}= \left(\begin{array}{c} {\widetilde d}_1
  \\ 0 \\ \vdots  \\ 0 \\   {\widetilde d}_1^{\prime}  \\ 0 \\ \vdots \\ 0  \end{array}\right). \label{firstsquark}\ee  }
According to our assumption  $x_{1}$ and $x_{2}$ are  two
distinct roots of $ m_{1}(x)=  0$, that is,
\be   \sum_k m_{1,k} \, x_{i}^{k-1} =0, \qquad  i=1,2.
\ee
Dividing the difference between these two equations by $x_{1}-x_{2}$,   one finds
\be    \sum_{k} \, m_{1,k}\,  \sum_{{\ell}=0}^{k-1} \, x_{1}^{\ell}\, x_{2}^{k-\ell - 1} =0.
\ee
This shows that  the color nondiagonal vacuum equation
Eq.(\ref{nondVEVs}) is indeed satisfied for $i=1$,  $c=1$,
$d=r_{1}+1$,  by  the  quark VEVs    Eq.(\ref{firstsquark}).

The generalization for more nonabelian factors, and for more
flavors getting VEVs in more than two color components, is
straightforward.

For simplicity of notations   and for definiteness, let us
restrict ourselves   in the following  to  the vacua with  only
two nonabelian  factors.    Their  multiplicity is given by the
combinatorial  factor
\be  {N_{f}  \choose {r_{1}}} \times  { N_{f} \choose r_{2}} \times \prod_{i=1}^{n} N_{i}
\ee
where the last factor corresponds to the Witten index of the pure
$\prod SU(N_{i}) \subset  U(N)$  theories.

Taking  the above considerations  into account,  the classical
VEVs  in our theory take, in the flavor symmetric limit,    the
form
\be
\label{phicondensate}
\langle\phi\rangle=\left(\begin{array}{ccccc}
v_1 {\bf 1}_{r_1}&&&&\\
&v_2 {\bf 1}_{r_2}&&&\\
&&a_1 {\bf 1}_{N_1}&&\\
&&&\ddots&\\
&&&&a_n {\bf 1}_{N_n}\\
\end{array}\right)\ ,
\ee
where
\be
\sum_{j=1}^{n} \, N_j+r_1+r_2=N,
\ee
and {\small  \be Q= \left(\begin{array}{cccccccc}
d_1 &  &  &  &  & & &         \\
& \ddots  &  &  &  & & &      \\
&  & d_{r_1} &  &  &  & &    \\
& e_1 &   &   &  &  & &       \\
&  & \ddots  &   &  &  & &    \\
&  &   &    e_{r_2} &  & &  &   \\
&  &   &   &  &  & &          \\
\end{array}\right) \ , \quad
 {\widetilde Q}= \left(\begin{array}{cccccccc}
{\widetilde d}_1 &  &  &  &  & & &         \\
& \ddots  &  &  &  & & &      \\
&  & {\widetilde d}_{r_1} &  &  &  & &    \\
& {\widetilde e}_1 &   &   &  &  & &       \\
&  & \ddots  &   &  &  & &    \\
&  &   &    {\widetilde e}_{r_2} &  & & &     \\
&  &   &   &  &  & &         \\
\end{array}\right) \ ,  \label{squark}
\ee  }
where
\be
d_c={\widetilde
d}_c=\sqrt{-\frac{W^{\prime}(v^{(1)})}{m^{\prime}(v^{(1)})}},
\qquad e_c={\widetilde
e}_c=\sqrt{-\frac{W^{\prime}(v^{(2)})}{m^{\prime}(v^{(2)})}}.
\ee
Note that classically, $0 \leq r_i \leq \text{min} [N_{f},N]$
apart from the obvious constraint  $\sum r_{i} \le N$.

As explained  above, the condensates $d_a$ and $e_b$ can share the
same flavor  and so there are ${N_f \choose r_1} {N_f  \choose
r_2}$ ways to choose Eq.(\ref{squark}).    We call $s $ the number
of ``superpositions'', that is,  the number of flavors that are
locked both to $v_1$ and $v_2$. The flavor symmetry is broken
then, in a $(r_{1}, r_{2})$ vacuum with $s$ superpositions,    as
 \be
U(N_f) \rightarrow U(r_1-s) \times U(s) \times U(r_2-s) \times U(N_f-r_1-r_2+s) \label{semicl}
\ee

The meson condensates  take  the form
\be
{\widetilde Q} Q =
\left(\begin{array}{cccccc}
-\frac{W^{\prime}(v^{(1)})} {m^{\prime}(v^{(1)})}  {\bf 1}_{r_1-s} &  &  &    &&\\
& -\left(\frac{W^{\prime}(v^{(1)})}{m^{\prime}(v^{(1)})} +\frac{W^{\prime}(v^{(2)})}{m^{\prime}(v^{(2)})}\right) {\bf 1}_{s}   &  &  &&  \\
&  & -\frac{W^{\prime}(v^{(2)})}{m^{\prime}(v^{(2)})} {\bf 1}_{r_2 - s}&   && \\
&  &   &  0 &&      \\
&&&& \ddots& \\
&&&&& 0 \\
\end{array}\right) \ ,
\label{meson}
\ee

The vacuum counting  becomes   even simpler  for  a quadratic
superpotential $W$,
\be
W^{\prime}(x)=g_0+g_1 x.    \label{linear}
\ee
In this case there is only one stationary point so the gauge group
is broken by the condensates to $U(N-r_1-r_2)$. The $SU$ part
confines and gives a Witten index $N-r_1-r_2$, therefore  the
total number of classical vacua is
\be
{\cal N}=\sum_{\substack{r_1,r_2=0,\dots N_f \\ r_1+r_2\leq N}}
(N-r_1-r_2) {N_f \choose r_1} {N_f \choose r_2}. \label{vac}
\ee
In the particular case $2N_f \leq N$ the second restriction on the
sum over $r_1$ and $r_2$ is absent, and the summation can be
performed to give a simple formula
\be
{\cal N} = (N- N_f) \, 2^{2N_f},
\ee
which is analogue of the  formula for the softly broken
${\cal N}=2$  SQCD valid for $N_{f}< N$ \cite{CKM},
\be {\cal N}_{SQCD} = (2\, N - N_{f}) \, 2^{N_{f}-1},\ee
and can be obtained from the latter by a formal replacement,
$N_{f} \to 2\, N_{f}$.

The semiclassical reasoning followed up to now is reliable  when
$|v_i| \gg \Lambda$, $|Q| \gg \lambda$,  and in this regime the
massless matter multiplets simply correspond to the first
$r_1+\ldots + r_{p} $ components of  the original quark
multiplets. When the parameters of the superpotential are such
that the vacuum expectation values of $\Phi$ and $Q, {\widetilde
Q}$    are of order of $\Lambda $ or smaller,  we expect these
massless multiplets to represent nonabelian magnetic monopoles.
The vacuum with the symmetry breaking Eq.(\ref{magnetic})  is more
appropriately seen as a vacuum in confinement phase, in which the
order parameters of confinement are various magnetic monopoles
carrying nonabelian charges
   \be   ({\underline {r_{1}} }, {\underline {\mathbf 1}}, {\underline {\mathbf 1}}, \ldots), \quad
    ({\underline {\mathbf 1}},  {\underline {r_{2}} }, {\underline {\mathbf 1}}, \ldots),
   \ee
    etc.

\section { Quantum  $(r_{1}, r_{2})$ Vacua \label{sec:quantum} }

When   $v_i$'s  are small, of order of $\Lambda$ or less, the
above semiclassical arguments are no longer  reliable, but by
varying continuously  the parameters   of  $m_{i}(\Phi)$  from
where the roots  $v_i$'s   are  all very large to the region where
they are small,  we expect that $\prod SU(r_{i})$ factors  remain
infrared free   or  superconformal,  as long as
\be   r_i   \le    \frac{ N_f }{ 2}, \quad  i=1,2,\ldots.
\ee
We conclude that in the theory Eq.(\ref{theory}) with $N_f$
quarks, in the limit
\be   W(\Phi) \to 0; \qquad    m_i(\Phi) \to m(\Phi), \label{limitns}
\ee
where   $m(\Phi)=0$  has at least two different  roots,
$\phi^*=v^{(1)}, v^{(2)},v^{(3)},\ldots,$ there must be vacua with
$\prod_i SU(r_i)$ effective gauge symmetry. If $|v_i| \le
\Lambda$  it must be a magnetic  theory  (the original $SU(N)$
interactions become strong).

We note that  since  the ultraviolet theory has no   massless
particles  having multiple nonabelian charges, such as
\be  ({\underline r}_1,  {\underline r}_2,  {\underline 1},  \ldots, ),
\ee
we do not expect massless monopoles with such   multiple  charges
to occur in the infrared either, in this theory.

The crucial information on the quantum system comes from  the
curve describing the  Coulomb branch of our
 ${\cal N}=1$ supersymmetric theory \cite{Kapustin}:
\be   y^{2} =    \prod_{i=1}^{N}  \, ( x - \phi_{i} )^{2}  -   \Lambda^{2 N- N_{f}} \, \det m(x)
= \prod_{i=1}^{N}  \, ( x - \phi_{i} )^{2}  -   \Lambda^{2 N-
N_{f}} \,\prod_{i}^{N_{f}}\, m_{i}(x), \label{Kap}
\ee
where  $m_{i}(x)$ is the function appearing in the superpotential
Eq.(\ref{theory}).   The curve is   valid for $\ell \, N_{f} < N$
($\ell$ being the order of the polynomial $m(\Phi)$).  For
\be  m_{i}(x)=    C \, (x - v_{i}^{(1)})\, ( x- v_{i}^{(2)}), \label{choice}
\ee
(so $\ell=2$)   the curve is
\be    y^{2} =    \prod_{i=1}^{N}  \, ( x - \phi_{i} )^{2}  - C^{N_{f}}   \, \Lambda^{2 N- N_{f}} \,\prod_{i=1}^{N_{f}} (x - v^{(1)}_{i})\, ( x- v^{(2)}_{i})
\label{curveKpn}
\ee
which is effectively equivalent to the curve of   the ${\cal
N}=2$, $SU(N)$ theory with   $2 \, N_{f}$ flavors.

The vacua with $U(r_{1}) \times U(r_{2}) \times U(1)^{N-r_{1}-r_{2}} $ low-energy
gauge symmetry arise at the point of QMS where the curve becomes
singular
\be   y^{2} =  (x-\alpha)^{2r_{1}}\, (x-\beta)^{2r_{2}} \,  [\,  \prod_{i=1}^{N-r_{1}-r_{2}}  \, ( x - \phi_{i} )^{2}
- C^{N_{f}}   \, \Lambda^{2 N- N_{f}} \,\prod_{i=1}^{N_{f}-2r_{1}}
(x - v_{i}^{(1)})\,\prod_{i=1}^{N_{f}-2r_{2}} ( x- v_{i}^{(2)})
\,], \label{effcurve}
\ee
with the factor in the square bracket factorized in maximum number of double factors.
These clearly occurs only in the flavor symmetric limit
$v_{i}^{(1)} \to v^{(1)}\equiv\alpha,\,$ $v_{i}^{(2)} \to
v^{(2)}\equiv\beta.\ $

As $r_{1}$ of $\phi_{i}$ can be equal to any $r_{1}$ of  $N_{f}$
$v_{i}^{(1)}$'s, and independently, $r_{2}$ of other
$\phi_{i}$'s    can be equal to any $r_{2}$ of  $N_{f}$
$v_{i}^{(2)}$'s,  there is a multiplicity
\be  { N_{f}\choose r_{1} }\,  {N_{f} \choose r_{2} }\,
\ee
of vacua which converge to the  $(r_{1}, r_{2})$ vacua in the flavor
symmetric limit. Actually, the ${\cal N}=1$  vacua exist at the
maximally abelian  singularity of the curve  Eq.(\ref{effcurve}),
which is, apart from  the factor  $(x- v^{(1)})^{2r_{1}}\,
(x-v^{(2)})^{2r_{2}}$, equivalent to that of ${\cal N}=2$
$SU(N-r_{1}- r_{2})$  gauge theory with massive matter. Therefore
the number of such  singularities is equal to $N-r_{1}- r_{2}$
\cite{SW1,SW2}. Collecting all the factors,  we find that  the
total multiplicity
\be {\cal N}=    { N_{f}\choose r_{1} }\,  {N_{f} \choose r_{2} }\, \, ( N-r_{1}- r_{2} )
\ee
which  coincides with  the semiclassical counting, Eq.(\ref{vac}),
for corresponding values of  $r_{1}, r_{2}$.

This last comment brings us to the subtle issue of correspondence
between classical and quantum $r$-vacua.  While classically, the
values of $r_{i}$ can reach $\text{min} (N, N_{f}), $   it is
evident from the above consideration that quantum vacua  exist
only for $r_{i} \le \frac{N_{f}}{2}.$   This nicely confirms and
generalizes the  importance of quantum effects in deciding which
nonabelian groups can survive in the infrared, emphasized
repeatedly by us.   At the same time,  we find that when the UV
theory contains an $SU(r_{i})$  factor supported by $N_{f}$
massless quarks,   $r_{i} >  N_{f}/2$,   so that the interactions
become strong at low energies,   such a sector is realized   in
the infrared  by the dual, magnetic $SU(N_{f}- r_{i})$ theory.
These nontrivial mappings between classical and quantum $r$-vacua
in softly broken ${\cal N}=2$  SQCD \cite{CKM} and in more general
models of the present  paper appear to explain nicely the origin
of Seiberg's duality.  We shall come back to this question in a
separate publication.

\section{Cachazo-Douglas-Seiberg-Witten Formulae } \label{sec:CDSW}

A further confirmation of our picture arises from the work by
Cachazo et al. \cite{CDSW}.
 In particular they solved for the  resolvents of the chiral operators
  \be M= {\widetilde Q} \, \frac{ 1 }{ z - \Phi} \, Q; \qquad  R(z)= - \frac{ 1}{ 32 \pi^{2}}\,\Tr \,
  \frac{W_{\alpha}\, W^{\alpha}  }{ z - \Phi}.
 \label{CRing}
 \ee
The main result is a set of the generalized anomaly equations \footnote{The notation
$\left[O(z)\right]_- $ stands for keeping only the negative powers in the
Laurent expansion of $O(z)$.}
\bqa && \left[W^{\prime}(z)\, R(z)\right]_-=R(z)^2,
\non \\
&&   \left[(M(z) \, m(z))_i^j\right]_- = R(z)\, \delta_i^j \, ; \qquad \left[(m(z)\, M(z))_i^j\right]_- = R(z)\, \delta_i^j.
\eea
$\det m(z)$ has $L= 2 \, N_{f}$ zeros $z_{i}$ (counted with their
multiplicity), which for the choice Eq.(\ref{choice}) can be
either $v_{1}$ or $v_{2}$ (in the $SU(N_{f})$ symmetric limit).
In fact,
  \be  m(z) =   \diag  \, [\, C\, (z- v_{i}^{(1)})\, (z- v_{i}^{(2)})\,],
  \ee
  \be   \frac{ 1 }{ m(z) } =  \left(\begin{array}{ccc} \frac{1 }{ C(z-v_1^{(1)})(z-v_1^{(2)})} &  &  \\ & \ddots &  \\ &  & \frac{1 }{ C(z-v_{N_{f}}^{(1)})(z-v_{N_{f}}^{(2)})}\end{array}\right).
\ee
where $  v_{i}^{(1)} \to v^{(1)}, \,\,   v_{i}^{(2)} \to v^{(2)}$
in the flavor symmetric limit.

The solution of the anomaly equations for $R(z)$ is
  \be  2 \, R(z) =  W^{{\prime}}(z) -   \sqrt{ W^{{\prime}}(z)^{2} + f(z) }
  \label{physical}
  \ee
where   $f(z)$ is directly related to the gaugino condensates in
the strong $\prod U(N_{i})$ sectors  \cite{CDSW}.  Thus  the zeros in $W^{\prime}(z)$  (denoted by $a_i$) which appear classically
as poles of  $\frac{1}{z- \Phi}$ are replaced by cuts in  a complex
plane $z$  by the quantum effects.  By
defining
\be  y=    W^{{\prime}}(z)  -  2 \, R(z),
\ee
one has a ${\cal N}=1$ curve
\be   y^{2} =      W^{{\prime}}(z)^{2} +   f(z) \label{mmcurve}
\ee
on this Riemannian  surface. The point of the construction of
\cite{CDSW}  lies in the fact that  various chiral condensates are
expressed elegantly in the form of integrals along  cycles on this
curve.

Taking the curve  Eq.(\ref{mmcurve})   as the
double cover of a  complex plane  with appropriate branch cuts,
the result of Eq.(\ref{physical}) refers to the ``physical''
(semiclassically visible) sheet; in the second sheet, the result
is ($\tilde{z}$ lies in the second sheet at the same value as $z$)
\be    W^{{\prime}}(z) =W^{{\prime}}({\tilde z}); \qquad
2\, R({\tilde z}) =  W^{\prime}({z})+\sqrt{ W^{{\prime}}(z)^{2} +
f(z) }.
\ee

 In the simplest model for $W$,
\be  W(\Phi)=  \mu  \, \Phi^{2}, \qquad   W(z)= \mu \, z^{2},
\ee
the only possible (classical) value for $a_{i}$ is zero; $f(z)$ is
a constant, $f=-8\mu S$, where
$$
S= -\frac{1}{ 32 \pi^2} \langle \Tr \, W_{\alpha}W^{\alpha}\rangle
$$
is the   VEV of the gaugino bilinear operator in the strong $SU(N-
r_{1}- r_{2})$ super Yang-Mills theory. The solution for $R$
is explicitly,
  \be  2 \, R(z) =  2\,  \mu \,  z -   \sqrt{ (2 \, \mu \, z)^{2} +   f  },
\qquad    2 \, R(\tilde{z}) =  2\,  \mu \,  z +   \sqrt{ (2 \, \mu
\, z)^{2} +   f  }.
\ee

The poles of $M$  in the classical theory, instead, remain poles
in the full quantum theory  \cite{CDSW}. We are interested in  a vacuum in
which $r_{1}+ r_{2}$   poles are in the physical  sheet:   $r_{1}$
poles near $v^{(1)}$  and  $r_{2}$ poles near $v^{(2)}$.  The
result of  \cite{CDSW}   for $M$  in this vacuum  is
\be   M(z) =  R(z) \frac{ 1 }{ m(z) }  -  \sum_{i=1}^{r_{1}+ r_{2}}  \frac{ R({\widetilde q}_{i})  }{  z - z_{i}} \,  \frac{ 1 }{ 2 \pi i} \,
 \oint_{z_{i}} \, \frac{ 1 }{ m(x)}
  \, dx  -  \sum_{j=1}^{ 2 N_{f}-r_{1}-r_{2}}  \frac{ R({q}_{j})  }{  z - z_{j}} \,  \frac{ 1 }{ 2 \pi i} \,  \oint_{z_{j}} \, \frac{ 1 }{ m(x)}
  \, dx.
  \label{symmbr}\ee
By definition the contour integrals must be done before the
$SU(N_{f})$  limit is taken, so
\be  \oint_{z_{i}} \,  \frac{1}{m(x)}\, dx = \left(\begin{array}{ccccc}   \ddots  & && & \\
&0 & &&   \\   & & \frac{1}{ m^{\prime}(z_{i})} & & \\  && &0 &    \\   &&&& \ddots   \end{array}\right),
\ee
where $m^{\prime}(v_{i})  = \pm \, C\, (v_{i}^{(1)}-v_i^{(2)})$,
the sign   depending  on whether  the  zero of $m(z)$  is  the one
near $v^{(1)}$ ($+$)    or near  $v^{(2)}$   ($-$). The
$N_{f}\times N_{f}$ flavor structure  is explicit in this formula:
for instance the first term reads
\be  [\, R(z) \frac{1}{m(z)} \,]_{ii}=   \frac{R(z)}{C(z-v_i^{(1)})(z-v_i^{(2)})}.
\ee
In the $SU(N_{f})$ symmetric limit, the first term is then
$\propto {\bf 1}_{N_{f }\times N_{f}}$; the second term of
Eq.(\ref{symmbr})  consists of $r_{1}$ terms  whose sum is
invariant under $U(r_{1}) \times U(N_{f}-r_{1})$ and $r_{2}$ terms
which form an  invariant under $U(r_{2}) \times U(N_{f}-r_{2})$.
The $r_{1}$ poles near $v^{(1)}$  can be related to any $r_{1}$
flavors out of $N_{f}$; analogously the $r_{2}$ poles near
$v^{(2)}$  can be associated  to any $r_{2}$ flavors out of
$N_{f}$: there is no restriction between the two subsets of
flavors.  It follows that the global symmetry is  broken  to
\be   U(N_{f}) \to  U(r_{1}-s) \times  U(r_{2}-s) \times  U(s)  \times  U(N_{f}- r_{1}-r_{2}+s),
\label{glsymbr}
\ee
in a vacuum of this type,
where $s$ is the number of ``overlapping'' flavors, {\it i.e.}, to
which both roots $v_{i}^{(1)} $ and $v_{i}^{(2)}$ appear as poles
in the first sheet. The result (\ref{glsymbr})  is perfectly
consistent with what was found  semiclassically,
Eq.(\ref{semicl}).

Actually a more precise correspondence between the semiclassical
and fully quantum mechanical results   is possible.  In order to
compare with the semiclassical result for the meson condensate,
Eq.(\ref{meson}),  it suffices to evaluate the   coefficient of
$\frac {1}{z}$  in the large $z$  expansion of the quantum formula
Eq.(\ref{symmbr})   (see  Eq.(\ref{CRing})).   $f$ is a constant
of order of   $\mu\Lambda^{3}$. We find
\be
{\widetilde Q} Q =
\left(\begin{array}{cccc}
A \, {\bf 1}_{r_1-s} &  &  &    \\
& B \, {\bf 1}_{s}   &  &    \\
&  & C\,  {\bf 1}_{r_2 - s}   & \\
&  &   &     D\, {\bf 1}_{N_{f}- r_{1}-r_2 + s}    \end{array}\right) \ ,
\label{Qmeson}
\ee
where \bqa A = - \frac  {R(\tilde v^{(1)}) } {
m^{\prime}(v^{(1)})} -  \frac  {R(v^{(2)}) } {
m^{\prime}(v^{(2)})} ;&& \quad B =  -   \frac  {R(\tilde v^{(1)})
} {  m^{\prime}(v^{(1)})}   - \frac  {R(\tilde v^{(2)}) } {
m^{\prime}(v^{(2)})}; \quad \non\\ C=  - \frac  {R (v^{(1)}) } {
m^{\prime}(v^{(1)})}   - \frac  {R(\tilde v^{(2)}) } {
m^{\prime}(v^{(2)})};&& \quad D=    -   \frac  {R (v^{(1)}) } {
m^{\prime}(v^{(1)})}   -        \frac  {R(v^{(2)}) } {
m^{\prime}(v^{(2)})}.
 \label{exact}
\eea

In the classical limit,  $f\to 0$, so
\be     R(v^{(1)}),   \,R(v^{(2)}) \to 0, \qquad    R(\tilde v^{(1)}) \to  W^{\prime}(v^{(1)}), \quad
 R(\tilde v^{(2)}) \to  W^{\prime}(v^{(2)}),
\ee
and  the quantum expression for the meson condensates
Eq.(\ref{Qmeson}) correctly reduces to Eq.(\ref{meson}).

\section{Classical vs  Quantum $r$-Vacua: Illustration
\label{sec:illust}}

Even though  we found a nice corresponding between the
semiclassical  and fully quantum mechanical  results, the precise
(vacuum by vacuum)  correspondence is slightly subtle, as the
ranges of $r_{i}$ are different in the two  regimes. In the
quantum  formulae,  the vacua are parametrized by $r_1 =
\min(N_f-r_1, r_1)$ and  $r_2 = \min(N_f-r_2, r_2)$. The curve
Eq.(\ref{Kap}) factorizes as
 \be   y^{2} =   (x-v_{1})^{2\, r_{1}}  \,   (x-v_{2})^{2\, r_{2}} \, (\ldots) \ .   \ee
The low-energy degrees of freedom, carrying nontrivial nonabelian charges,  are given in Table \ref{low}.\begin{table}[ht]
\begin{center}
\begin{tabular}{cccccc}
 $U(N_f)$ & $SU(r_1)$ & $U(1)$ & $SU(r_2)$ & $U(1)$ & $U(1)^{N-r_1-r_2}$ \\
\hline
  $\underline{N}_f$ & $\underline{r}_1$ & $1$ & $\underline{1}$ & $0$ & $0$\\
  $\underline{N}_f$ & $\underline{1}$ & $0$ & $\underline{r}_2$ & $1$ & $0$\\
\hline
\end{tabular}
\end{center}
\caption{\footnotesize Massless dual-quarks in $r_1, r_2$-vacua.}
\label{low}
\end{table}

Clearly, such a stricter condition  for  $r_{i}$  for the quantum
theory reflects the renormalization effects due to which  only for
   $r_{i}$ less than $N_{f}/2$  these nonabelian interactions  remain non asymptotically free and  can survive
   as low-energy gauge symmetries.
We find the following correspondence between the classical ($r_{1}, r_{2}$)   vacua  and the quantum  ($r_{1}, r_{2}$)     vacua:
\be
\left.\begin{tabular}{cc}
$r_1$  & $r_2$ \\
$r_1$ & $N_f-r_2$ \\
$N_f-r_1$ & $r_2$ \\
$N_f - r_1$ & $N_f - r_2$\\
\end{tabular}\right\} \rightarrow \begin{tabular}{cc}
$r_1$ & $r_2$ \\
\end{tabular}
\ee
and the total matching of the number of vacua in  the two regimes
must take  into account of these rearrangements of $r_1, r_2$.
(This occurs also in the simpler case of the softly broken ${\cal
N}=2$ SQCD  of Carlino et.al.\cite{CKM})


In order to check the whole discussion  and to illustrate  some of
the results found,   we have performed a numerical study   in the
simplest nontrivial models.
 Figure \ref{corr1} shows the situation for  $U(3)$
theory  with nearly degenerate  $N_{f}=4$ quark flavors and with a
linear function  $m(\Phi)$.  This corresponds basically to the
seventeen vacua of the $SU(3)$ theory  studied earlier in
the context of softly broken ${\cal N}=2$ SQCD \cite{CKM} (plus four vacua of $r=3$
due to the fact that here we consider $U(3)$). The second
example, Figure \ref{corr2},  refers to a $U(3)$ gauge theory with
$N_{f}=2$, but  with a quadratic   function $m(\Phi)$, illustrates well  the situations studied   in the present  paper.  We start from the curve Eq.(\ref{Kap})
in the flavor-symmetric limit
\be   y^{2} =  G(x)=  \prod_{i=1}^{N}  \, ( x - \phi_{i} )^{2}  -   \Lambda^{2 N- N_{f}} \, \det m(x)
= \prod_{i=1}^{N}  \, ( x - \phi_{i} )^{2}  -   \Lambda^{2 N-
N_{f}} \,\, m(x)^{N_{f}}, \label{Kapp}
\ee
and apply the factorization equation of \cite{CDSW,Feng}
\bqa G(x)= F(x) H^2(x) \label{feq1} \\
W^{\prime}(x)^2+f(z) =F(x)Q^2(x) \label{feq2}
\eea
In our model we take
\be     W(x)= \mu \, x^{2},
\ee
so $\deg Q(x) \in \{0,1\}$. In particular $\deg Q=0$ in a vacuum
smoothly connected with a classical vacuum in which $\Phi$ is
completely higgsed, $\langle\Phi\rangle=\diag(z_1,\ldots,z_{N})$,
whereas $\deg Q=1$ corresponds to a classical VEV for $\Phi$ of
the type $\langle \Phi\rangle=\diag(z_1,\ldots,z_{r},0,\ldots,0)$,
where some eigenvalues are zeros of the adjoint superpotential
$W(\Phi)$. In other words $\deg Q=0$ or $1$ if the matrix model
curve Eq.(\ref{mmcurve}) is degenerate or not respectively.

We must require the vanishing of the discriminant of the curve
$G(x)$, that is the resultant of $G(x)$ and its first derivative:
\be {\cal R} \left(G(x),\frac{dG(x)}{ dx} \right)=0.
\ee
This guarantees the presence of a double zero in $G(x)$.
\begin{description}
    \item[A.] \emph{$U(3)$ $N_f=4$ with linear $m(\Phi)$ (see Fig.\ref{corr1}).}\\
\begin{figure}[ht]
\begin{center}
\leavevmode
\epsfxsize 12   cm
\epsffile{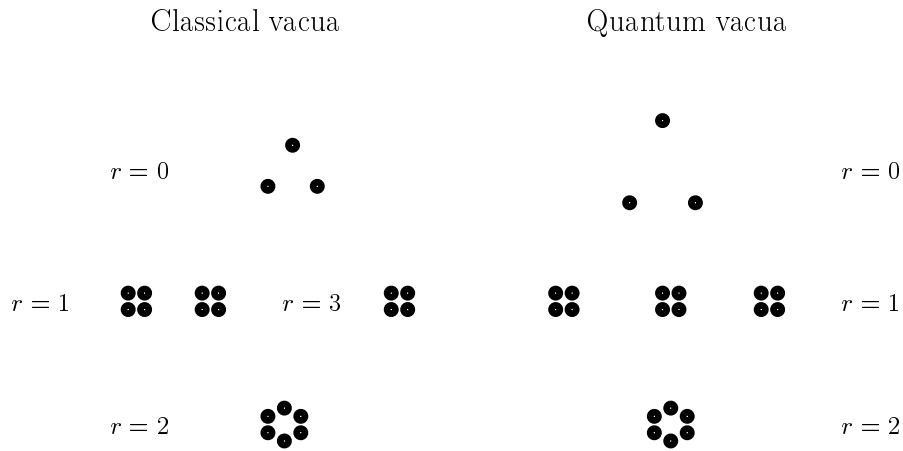}
\end{center}
\caption{\footnotesize Correspondence between the classical and
quantum vacua in the $U(3)$ theory with $4$ flavors at degenerate
mass.    } \label{corr1}
\end{figure}
    \be y^{2} =  G(x)=  \prod_{i=1}^{3}  \, ( x - \phi_{i} )^{2}  -   \Lambda^{2} \, (x-
    m)^4 \label{u3linear}
    \ee

    As explained above quantum vacua are labeled by an integer
    $r$ that ranges from $0$ to
    $N_f/2=2$.
    \begin{description}
    \item[(i)] $r=2$.\\
    After extracting a factor $(x-m)^4$ from $G(x)$ we are left
    with a reduced curve
    \be \tilde{y}^2=\tilde{G}(x)=(x-a)^2-\Lambda^{2}
    \ee
    Imposing Eqs.(\ref{feq1}),(\ref{feq2}) we can fix $a=0$ in order to get two
    opposite zeros for the reduced curve (the only possible choice is indeed $\deg Q=0$). So there is only  one
    solution, that is only one quantum vacuum with $r=2$;
    its multiplicity is $_{N_f}C_r=6$, as we can check performing a
    quark mass perturbation on the curve. Then we have actually a
    sextet of $r=2$ quantum vacua. These correspond exactly to the
    six classical vacua with $r=2$.

    \item[(ii)] $r=1$.\\
    We extract a factor $(x-m)^2$; the reduced curve is
    \bqa \tilde{y}^2 &=& \tilde{G}(x)=(x^2-ax-b)^2-\Lambda^{2}(x-m)^2
    \eea
    The discriminant of $\tilde{G}(x)$ vanishes on particular
    (complex) 1-dimensional submanifolds of the moduli space parametrized by $(a,b)$.
    Clearly we must exclude those where
    $\tilde{G}(x)=(x-m)^2\ldots$, because it belongs to the case
    $r=2$ again. Afterwards we have two possibilities; looking at
    Eq.(\ref{feq2}) we can choose $\deg Q=0$ and adjust the remaining
    free parameter to get one more double zero of $\tilde{G}(x)$
    or $\deg Q =1$ and let $\tilde{G}(x)$ have two opposite zeros.
    We recover one solution for the first choice (this is what is called baryonic root in \cite{CKM}) and two solutions
    for the second. All these vacua have multiplicity $_{N_f}C_r=4$
    and they correspond to classical vacua with $r=1,3$.

    \item[(iii)] $r=0$.\\
    We work with the full curve Eq.(\ref{u3linear}). Requiring the
    vanishing of the discriminant and avoiding the set of
    solutions that lead to the form $G(x)=(x-m)^2\ldots$ (it would
    belong to the above cases) we find 3 solutions by imposing
    Eq.(\ref{feq2}) with $\deg Q=0$. All these vacua have multiplicity
    $_{N_f}C_r=1$ and they correspond to classical $r=0$ vacua.

    \end{description}

    \item[B.] \emph{$U(3)$ $N_f=2$ with quadratic $m(\Phi)$ (see Fig.\ref{corr2}).}\\
\begin{figure}[ht]
\begin{center}
\leavevmode
\epsfxsize 13   cm
\epsffile{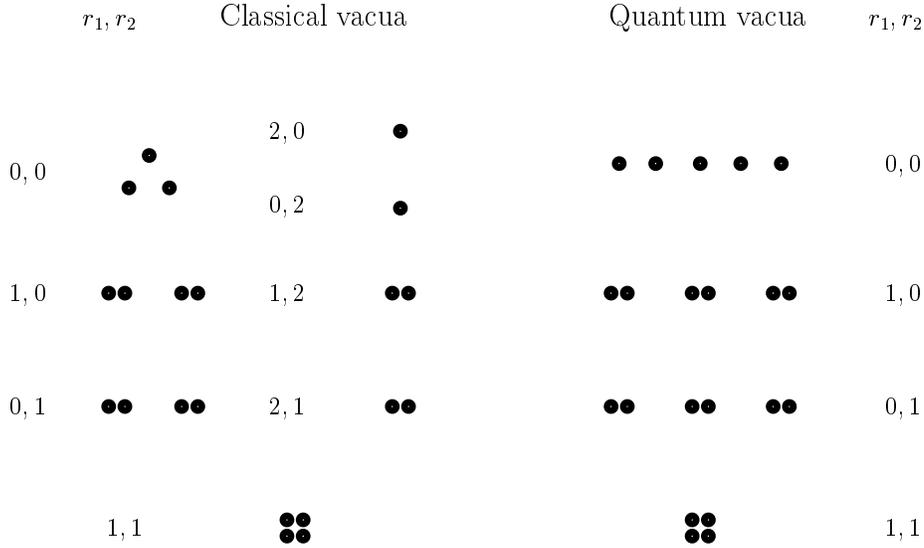}
\end{center}
\caption{\footnotesize  Correspondence between the classical and
quantum vacua in the $U(3)$ theory with $2$ flavors and quadratic
$m(x)$ or equivalently $4$ flavors with two different masses.}
\label{corr2}
\end{figure}
    \be   y^{2} = G(x)=   \prod_{i=1}^{3}  \, ( x - \phi_{i} )^{2}  - C^{2}   \, \Lambda^{2} \, (x - v^{(1)})^2\, ( x-
    v^{(2)})^2
    \label{u3quadratic}
    \ee
    Here we work with two ``quantum'' indices $r_1,r_2\in\{0,1\}$.

    \begin{description}
        \item[(i)] $(r_1,r_2)=(1,1)$.\\
        The curve Eq.(\ref{u3quadratic}) contains a factor $(x - v^{(1)})^2\, ( x-
        v^{(2)})^2$ so we are left with
        \be \tilde{y}^2=\tilde{G}(x)=(x-a)^2-C^2\Lambda^{2}
        \ee
        There is only one solution for $a$ just as in A.(i), but
        now the multiplicity of the vacuum is
        $_{N_f}C_{r_1}\times\,
        _{N_f}C_{r_2}=2\times 2=4$. This is what expected from the
        counting of classical $(r_1,r_2)=(1,1)$.

        \item[(ii)] $(r_1,r_2)=(0,1)$.\\
        The reduced curve is
        \be \tilde{y}^2 = \tilde{G}(x)=(x^2-ax-b)^2-C^2\Lambda^{2}(x-v^{(1)})^2.
        \ee
         As in
         A.(ii) we can have $\deg Q=0$ or $1$; we obtain
         respectively one and two solutions for the parameters
         $(a,b)$. The multiplicity of this three vacua is
         $_{N_f}C_{r_1}\times\,
        _{N_f}C_{r_2}=1\times 2=2$, so we recover the desired
        number coming from classical vacua with
        $(r_1,r_2)=(0,1),(2,1)$.

        \item[(iii)] $(r_1,r_2)=(1,0)$.\\
        This case is exactly as in B.(ii) with $v^{(1)}\leftrightarrow
        v^{(2)}$ and $r_1 \leftrightarrow r_2$.

        \item[(iv)] $(r_1,r_2)=(0,0)$.\\
        Studying the full curve Eq.(\ref{u3quadratic}) and excluding
        solutions carrying a factor $(x-v^{(1)})^2$ or $(x-v^{(2)})^2$ in
        $G(x)$, we find five solutions to the factorization
        equations Eqs.(\ref{feq1}),(\ref{feq2}) ($\deg Q=0$). Their
        multiplicity is $_{N_f}C_{r_1}\times
        _{N_f}C_{r_2}=1$. This allows us to recover the right
        number of vacua corresponding to classical
        $(r_1,r_2)=(0,0)$ (multiplicity 3) and
        $(r_1,r_2)=(2,0),(0,2)$ (multiplicity 1).

    \end{description}

\end{description}

\section{Conclusion}

In this paper we have shown that under some nontrivial conditions
a $U(N)$ gauge theory with ${\cal N}=1$ supersymmetry  is realized
dynamically    at low energies  as  an effective  multiply
nonabelian gauge system
\be   U(r_{1}) \times U(r_{2})\times  \ldots  \times  \prod U(1),
\label{existence}\ee
with  massless particles    having charges $({\underline
{r_{1}}}, {\underline {\bf 1}},\ldots)$, $( {\underline {\bf 1}},
{\underline {r_{2}}}, {\underline {\bf 1}}, \ldots)$, ...
In the fully quantum situation  discussed in Sections
\ref{sec:quantum}, \ref{sec:CDSW}, \ref{sec:illust},    these
refer to magnetic particles carrying  nonabelian charges so the
 gauge symmetry breaking induced by the superpotential $W(\Phi)$  (dual Higgs mechanism)  describes
a {\it  nonabelian  dual superconductor of a more general type}  than studied earlier.

We believe that the significance  of our work lies not in a
particular model considered or its possible applications, but in having given an existence proof of
 $U(N)$  systems which are  realized at low energies as a magnetic gauge theory  with multiple nonabelian gauge group factors.  We found the conditions under which  this type of  vacua are realized.
 As a bonus, an intriguing  correspondence between classical and quantum
 $(r_{1}, r_{2}, \ldots)$ vacua  was found,  generalizing an analogous phenomenon in the
 standard softly broken  ${\cal N}=2$  SQCD.
 We think that our findings constitute a small but useful step towards a more ambitious goal of
achieving a complete classification  of  possible confining systems in 4D gauge
theories, or finding the dynamical characterization of each of them.

\section* {Acknowledgement}

The authors  thank  Roberto Auzzi, Jarah  Evslin, Luca Ferretti and Naoto
Yokoi for comments and discussions.

\end{document}